\documentclass[aps,pra,reprint,superscriptaddress]{revtex4-2}
\usepackage{graphicx}
\usepackage{color}
\usepackage{amsmath,amssymb}
\usepackage{bm}
\usepackage{upgreek}
\usepackage{xspace}
\usepackage[colorlinks,urlcolor=blue,citecolor=blue,linkcolor=blue]{hyperref}

\usepackage{bbold}

\usepackage[english]{babel}

\begin{document}

\title{Measurement of a quantum system using spin-mechanical conversion}
\author{A.~A.~Wood}
\email{alexander.wood@unimelb.edu.au}
\affiliation{School of Physics, University of Melbourne, Victoria 3010, Australia}
\author{D. S. Rice}
\affiliation{School of Physics, University of Melbourne, Victoria 3010, Australia}
\author{T. Xie}
\affiliation{School of Physics, University of Melbourne, Victoria 3010, Australia}
\author{F. H. Cassells}
\affiliation{School of Physics, University of Melbourne, Victoria 3010, Australia}
\author{R. M. Goldblatt}
\affiliation{School of Physics, University of Melbourne, Victoria 3010, Australia}
\author{T. Delord}
\affiliation{Department of Physics, City College of the City University of New York, New York, New York 10031, USA}
\author{G. H\'etet}
\affiliation{Laboratoire De Physique de l’\'Ecole Normale Sup\'erieure, \'Ecole Normale Sup\'erieure, PSL Research University, CNRS,
Sorbonne Universit\'e, Universit\'e de Paris, 24 rue Lhomond, 75231 Paris Cedex 05, France}
\author{A. M. Martin}
\affiliation{School of Physics, University of Melbourne, Victoria 3010, Australia}

\date{\today}
\begin{abstract}
Levitated macroscopic particles exhibiting quantum mechanical effects are garnering increased attention as a means for precision sensing and testing quantum mechanics. Defects in diamond, such as the nitrogen-vacancy (NV) centre possess optically-addressable spins with long coherence times at room temperature and offer an intriguing system to examine quantum spin dynamics coupled to a macroscopic classical particle. In this work, we convert the outcome of a quantum measurement on an ensemble of spins into a macroscopic rotation of the host particle via spin-mechanical coupling. Following a sequence of green laser and microwave control pulses, spin-mechanical coupling between the final qubit spin state and the host particle -- an electrically-levitated diamond -- exerts a torque on the particle that deflects a weak near-infra-red laser beam. We measure spin readout contrast in excess of 70\%, and demonstrate pulsed mechanical detection of coherent Rabi oscillations, spin-echo interferometry and $T_1$-induced relaxation. We directly measure with temporal resolution the particle reorientation from a 60\,attonewton-metre spin torque induced by flipping the spins. Our results open up interesting new opportunities for levitated spin-mechanical systems using pulsed control, from improved sensing to the prospect of realising macroscopic quantum superposition states.   
\end{abstract}
\maketitle
{\bf Introduction.} 
The connection between magnetism and physical rotation has been established for more than a century~\cite{einstein_experimenteller_1915, barnett_magnetization_1915}, long before spin was established as a quantum phenomenon. Every quantum operation performed on a spin qubit imparts a tiny mechanical force by virtue of conservation of angular momentum, and for solid-state spins these forces can ultimately entangle the spin with the mechanical motion of the host substrate. Levitation of spin-hosting nano- and microparticles has been intensively pursued in the quest to realise quantum mechanical systems coupled to mechanical oscillators~\cite{yin_hybrid_2015, pettit_coherent_2017, delord_electron_2017, conangla_motion_2018,obrien_magneto-mechanical_2019, millen_optomechanics_2020, huillery_spin_2020,gonzalez-ballestero_levitodynamics_2021,perdriat_spin-mechanics_2021, rusconi_spin-controlled_2022, fuwa_ferromagnetic_2023, wachter_gyroscopically_2025}. Diamond has received specific attention, owing to the presence of optically-addressable, long-lived spin qubits such as the NV centre~\cite{doherty_nitrogen-vacancy_2013}, which can exhibit spin coherence times of $T_2^\ast, T_2 > 1\,$ms in bulk samples~\cite{balasubramanian_ultralong_2009, herbschleb_ultra-long_2019} and a $T_2$ of hundreds of microseconds in milled CVD nanodiamonds~\cite{wood_long_2022}. The embedded spin serves to couple quantum spin dynamics to the motion of the particle, yielding a hybrid spin-mechanical system with fundamental applications such as testing the length scale on which quantum superpositions can exist~\cite{romero-isart_large_2011, scala_matter-wave_2013,yin_large_2013,wan_free_2016} and searching for possible signatures of quantum gravity~\cite{bose_spin_2017, bose_spin-based_2025}. While optical trapping of diamonds is challenging due to the absorption of trapping light causing uncontrolled heating in vacuum~\cite{rahman_burning_2016}, electrical traps such as the Paul trap have been employed in several experimental demonstrations of spin-mechanical control~\cite{delord_strong_2017, pellet-mary_magnetic-torque_2021}, most notably trapping of single-NV hosting nanodiamonds~\cite{conangla_motion_2018}, spin-cooling the librational motion of a trapped microdiamond~\cite{delord_spin-cooling_2020} and quantum control at rotation speeds of 20\,MHz~\cite{jin_quantum_2024}. 

\begin{figure*}
	\centering
		\includegraphics[width = \textwidth]{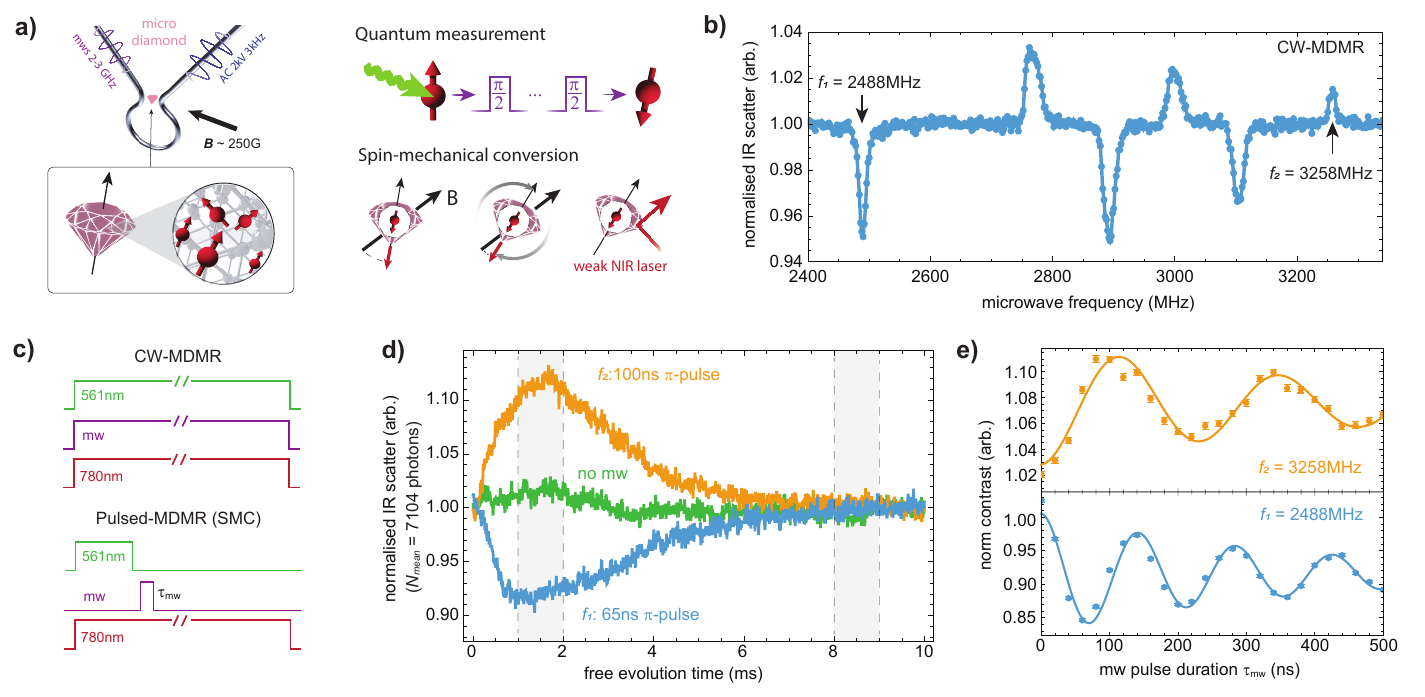}
	\caption{\textbf{Measurement via spin-mechanical conversion.} a) Experimental schematic: a microdiamond with $\sim 10^8$ NV centres inside is levitated in a Paul trap. Quantum measurement: green light optically pumps NV centres inside the diamond into the $m_S = 0$ spin state, and a sequence of microwave pulses executes a quantum measurement on the spin ensemble. Spin-mechanical conversion: the projected outcome of the spin measurement results in a magnetisation that applies a torque to the levitated crystal. The crystal rotation is then measured by detecting scattered weak NIR light. b) cw-MDMR spectrum of scattered NIR light measured with continuous application of green, NIR and swept microwaves (sequence shown in (c), top). Only six of the eight resonances are visible here. We identify two resonances with opposite contrasts at $f_{1,2} = 2488, 3258\,$MHz corresponding to $m_S = 0\rightarrow\pm1$ spin transitions. c), bottom: pulsed-MDMR sequence: a 200-500\,$\upmu$s 100\,$\upmu$W green laser pulse optically pumps the NV centres, and a microwave pulse flips the target spin transition. Continuous application of non-perturbative NIR light facilitates readout of the motional transient induced by the mw pulse. d) Normalised NIR scatter for no mw pulse and $\pi$-pulses on the $f_{1,2}$ transitions. Grey dashed regions denote integration regions used in e), which shows mw Rabi oscillations on each spin transition and readout contrasts of 10-15\%. Error bars deduced from photon counting statistics.}
	\label{fig:fig1}
\end{figure*}

A key technique introduced by Delord \emph{et al.} \cite{delord_spin-cooling_2020} was mechanically-detected magnetic resonance (MDMR), where spin-mechanical coupling between a large ensemble ($\sim10^9$) of NV centres and the librational motion of the host microdiamond yields a mechanical reorientation of the particle due to the induced NV magnetisation. The particle motion is subsequently detected by monitoring scattered light, rather than NV photoluminescence. In Ref~\cite{delord_spin-cooling_2020}, the green light used to control the NV colour centres was also used for motional detection. Consequently, spin-dependent reorientation of the trapped particle competed with repolarisation induced by the laser, precluding mechanical measurement of pure spin dynamics and extensions into the regime where coherent spin dynamics impact the motion of the particle.

Here, we realise non-perturbative, mechanical detection of an ensemble of quantum spins prepared by a sequence of control pulses in a levitated microparticle. The outcome of a standard sequence of spin polarisation and manipulation controls applied to the NV spins is stored as a magnetisation, which spin-mechanical conversion (SMC) then transduces into mechanical rotation of the host microdiamond. We continuously detect the particle motion with a weak near-infra red laser, which has a negligible effect on both the spin and motion of the particle, allowing for longer measurement times, greater particle movement and larger optical deflections. We show that SMC serves as a high-fidelity readout method for pulsed spin manipulation techniques, demonstrating high-contrast, mechanically-detected Rabi oscillations and spin-echo interferometry, with the latter effectively transducing the quantum phase acquired by the spin ensemble into a mechanical, macroscopic particle reorientation. We quantify the effects of spin relaxation on the torque imparted by the spins, and directly measure the particle reorientation --and hence torque -- using mechanically-detected vector magnetometry. Our work opens new sensing opportunities for levitated spin systems, enables a vast array of coherence-extending spin control sequences to be employed and advances the prospect of probing the quantum dynamics of an ensemble of quantum spins mechanically coupled to a classical oscillator. 

{\bf Results.} 
Our experiment~(Fig. \ref{fig:fig1}(a)) is based on Ref. \cite{delord_spin-cooling_2020} and described fully in Methods and the accompanying Supplementary Information. Briefly, a $10\,\upmu$m microdiamond trapped in a Paul trap is illuminated by green (561\,nm) and near IR light (780\,nm), for NV spin pumping and motional detection, respectively. Near IR light (NIR, 747\,nm) was also used to monitor motion induced by NV magnetism in Ref. \cite{perdriat_angle_2022} without impacting the NV spin state. Microwaves are applied through the Paul trap electrodes to manipulate the NV spin, and scattered NIR light is collected to monitor the motion of the particle. The Paul trap voltage and AC frequency are adjusted so that the trapped diamond librates about an axis rather than rotates~\cite{perdriat_rotational_2024}. We first perform a standard cw-MDMR experiment in the presence of a magnetic field to identify resonant microwave transitions. Figure~\ref{fig:fig1}(b) shows a cw-MDMR spectrum of scattered NIR light acquired with continuous application of green light and microwaves, revealing the characteristic high-contrast features of three of the four possible NV orientation classes, with opposite measurement contrast corresponding to microwave induced $m_S = 0\rightarrow\pm1$ spin transitions. We consider two transitions corresponding to the same orientation class and opposite spin polarity located at $f_{1,2} = 2488, 3258\,$MHz. In cw-MDMR, Figure~\ref{fig:fig1}(c, top), spin-induced reorientation of the particle is countered by green laser pumping populating the non-magnetic $m_S = 0$ spin state, limiting the particle angular displacement and hence measurement contrast. 

{\bf SMC readout.} Pulsing the green light and then applying a resonant microwave pulse (Fig.~\ref{fig:fig1}(c, bottom)) allows reorientation of the particle due to the spin magnetisation alone to be detected. Monitoring the scattered NIR light over 10\,ms Fig.~\ref{fig:fig1}(d), we observe unambiguous transients in the collected photon time traces corresponding to particle reorientation in opposite directions. The characteristic shape of the transient can be intuitively understood (see Theory and Supplementary Information) first as a spin-induced torque generated by the magnetic spin state population acting to orient the diamond particle so that the magnetisation aligns closer to the applied magnetic field~\cite{delord_spin-mechanics_2019, perdriat_angular_2023} over the first 1\,ms of evolution. Subsequently, relaxation parametrised by $T_1$ erodes the spin polarisation and attenuates the spin torque before the trap restoring force re-establishes the prior particle orientation. We then compare photon counts in each of these two regimes (approx. 1\,ms and 8\,ms respectively) to define a SMC contrast $C$, and vary the duration of the applied microwave pulse on both $f_1$ and $f_2$ transitions, tracing out Rabi oscillations with opposite contrasts due to the different spin torque signs as shown in Fig. \ref{fig:fig1}(e).    

{\bf Detection with SMC.} Having introduced and demonstrated SMC, we now compare it to standard photoluminescence (PL) readout. The measurement contrast obtained in Fig. \ref{fig:fig1}(e) already exceeds the maximum PL contrast attainable for a single orientation class in an ensemble NV sample of $8\,\%$, a limit only achievable in the absence of background fluorescence from other defects, chiefly NV$^0$. The ultimate limit on MDMR contrast is set by how effectively spin-dependent scattered light from the particle can be isolated at the detector, and can in principle be unity. In our experiment, the reflected light from the particle is focused onto a small aperture in a confocal arrangement, a 50\,$\upmu$m optical fiber. Libration and motion of the diamond leads to variations in the collection efficiency, transducing motion into measured photon number. We found that this arrangement is simultaneously appropriate for PL collection while yielding high MDMR contrasts in excess of previous demonstrations, where local variations in an unfocused speckle pattern of reflected light were sampled by a single-mode fiber~\cite{delord_spin-cooling_2020}. 

\begin{figure}
	\centering
		\includegraphics[width = \columnwidth]{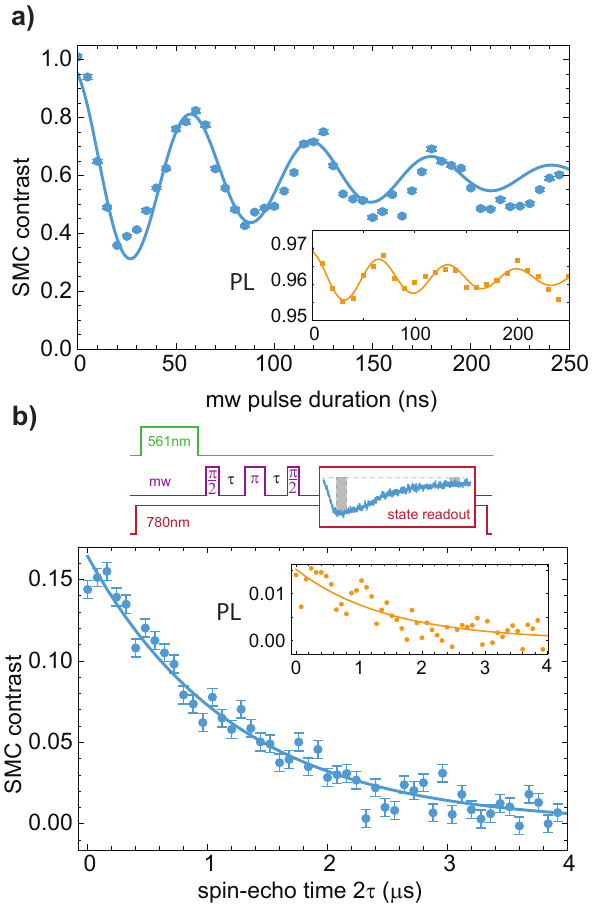}
	\caption{\textbf{SMC vs PL.} a) SMC readout of Rabi oscillations averaged for 60\,seconds, exhibiting 73(6)\,\% measurement contrast, 50 times greater than Rabi oscillations detected via conventional PL measurements in 30 mins of averaging (inset, in orange). b), top: pulse sequence for mechanically-detected spin-echo, and bottom: results (10\,min averaging), again compared with standard PL readout (inset, 33\,min averaging). Error bars deduced from photon counting statistics.}
	\label{fig:fig2}
\end{figure}

In Fig. \ref{fig:fig2}, we compare SMC readout to standard PL. The PL contrast we measure, typically $1$-$2\,\%$, is well below the ideal limit of 8\,\%~\cite{rondin_magnetometry_2014}, a situation common in very-high N density irradiated diamond due to a significant NV$^0$ population~\cite{tallaire_high_2020}. The SMC contrast, however, is unaffected by the NV$^0$ population as charge transfer between the two species does not lead to a measurable torque, and in the example given in Fig. \ref{fig:fig2}(a) is 73(6)\,\% after 60\,sec of measurement time, close to double the maximum PL contrast of 30-40\% achievable for a single NV centre in ultrapure diamond with charge state optimisation~\cite{barry_sensitivity_2020, song_pulse-width-induced_2020, wirtitsch_exploiting_2023} and 50 times better than standard PL Rabi measurements (inset) after 30\,min of averaging time. Both the PL and scattered NIR are attenuated by a factor of 10$^4$ to prevent damage to our single-photon counting detector. Photon count rates in both detection modalities are between 1-5\,Mcts/s, making direct comparison possible. 

The impressive Rabi contrast offered by SMC carries over to more sophisticated measurements, and in Fig. \ref{fig:fig2}(b) we compare mechanically-detected spin-echo to that achievable with PL measurements. In spin-echo, the quantum phase accumulated by the spins during the free evolution time is first projected into a a population distribution, which in turn becomes particle orientation during the free evolution time. As such, our measurement constitutes mechanical detection of quantum information, a first for levitated spin systems. We note here that these measurements were conducted for two separate diamonds, and in the case of the PL measurement the diamond was detrapped and electrically adhered to an electrode, rendering it stationary and preventing contribution of mechanical noise to the PL readout.

{\bf Relaxation and mechanical readout.} Under continuous green laser illumination, the MDMR signal arises from a competition of laser-induced population of the $m_S = 0$ state and microwave-induced occupation of magnetic $m_S = \pm1$ states. As such, intrinsic spin coherence plays a negligible role beyond broadening the microwave transition frequency. While both $T_2 = 1.22(5)\,\upmu$s and $T_2^\ast<100\,$ns are far too short to have an impact on the librational dynamics of the particle, we can easily observe the effects of $T_1$ relaxation on the particle dynamics with SMC. As described earlier, $T_1$ serves to attenuate the spin torque applied by the polarised NV spins for increasing times, leading to the inflection of the scattered NIR light vs time traces near 1\,ms. 

\begin{figure}
	\centering
		\includegraphics[width = \columnwidth]{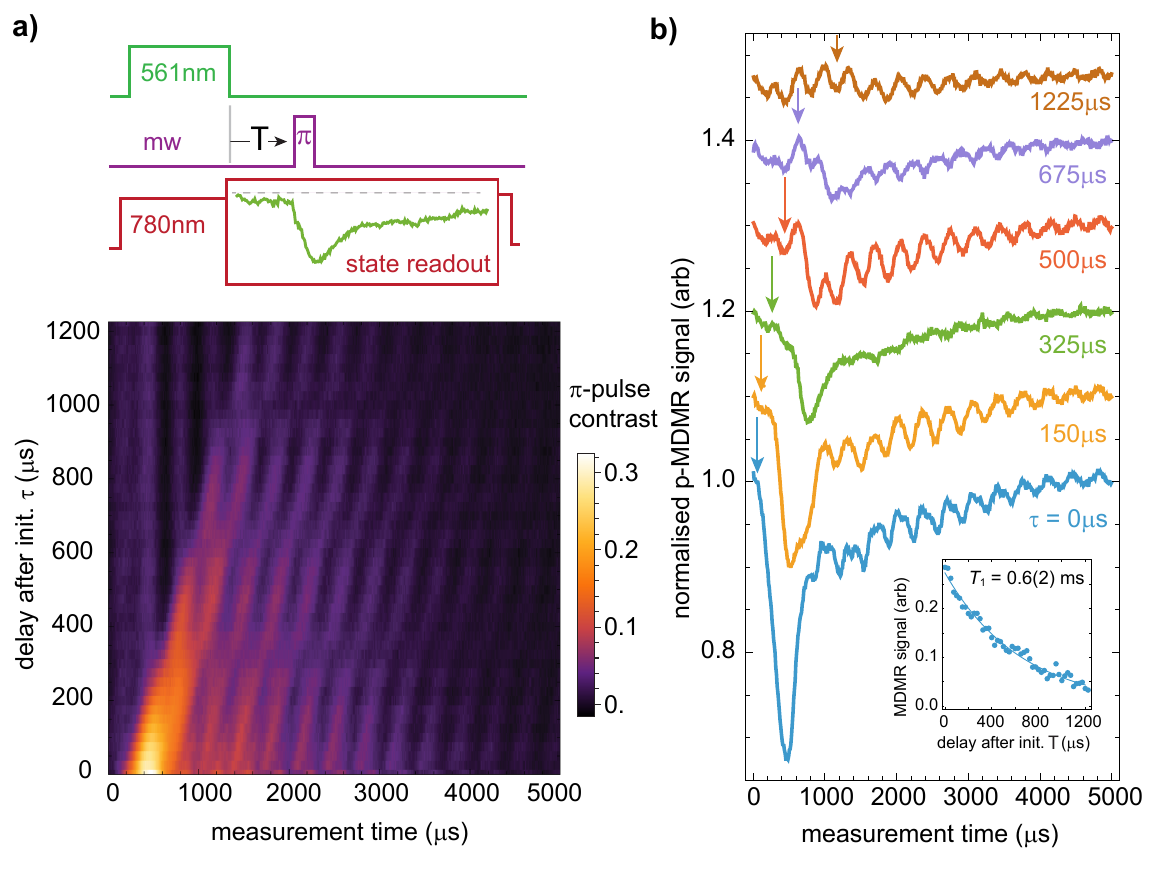}
	\caption{\textbf{$T_1$ measurement via mechanical readout.} a) Pulse sequence (top) and motional transient data vs delay time (bottom). The decay of the response is due to $T_1$ induced depolarisation, which attenuates the spin torque. Micromotion due to both the initial laser polarisation pulse and subsequent microwave $\pi$ pulse at the trap AC frequency is clearly evident, as well as mutual cancellation at multiples of $1/f_\text{trap}$. b) Detail, showing time traces vertically offset for clarity. The integrated signal response $F$ is plotted inset, from which $T_1$ may be determined.}
	\label{fig:fig3}
\end{figure}

We measure $T_1$ by first initialising the spin ensemble into the $m_S = 0$ state with a green laser pulse and then waiting some duration $T$ before applying a microwave $\pi$-pulse, as shown in Figure \ref{fig:fig3}(a). As the delay time $T$ increases, the amplitude of the scattered NIR transient reduces due to the depolarising effect of $T_1$. A measure of the signal intensity, $F = \int_T^{5\,\text{ms}}S(t)dt$ where $S(t)$ is the scattered NIR time trace is plotted vs $\tau$ in the inset of Fig. \ref{fig:fig3}(b), which yields $T_1 = 0.6(2)\,$ms. In this data, we observe an additional effect due to the perturbative effect of both the green laser pulse and the microwave pulse: induced micromotion at the trap AC drive frequency, $2895\,$Hz for this diamond. Radiation pressure induced by the green laser pulse displaces the particle in the trap and induces micromotion, a well-known phenomenon in ion traps. However, this motion which appears as vertical lines in Fig. \ref{fig:fig3}(a), can be cancelled by micromotion induced by the spin torque reorientation of the particle. The spin torque displaces the particle's orientation from the trap equilibrium point and also induces micromotion in the librational degree of freedom, both translational motion and orientation changes are detected in our experiment. We observe induced micromotion-free traces at intervals of $1/f_\text{trap}$ (Fig. \ref{fig:fig3}(b)). The effects of intrinsic particle micromotion, which can be the dominant particle motion, are detailed further in the Supplementary Information.

\begin{figure}
	\centering
		\includegraphics[width = \columnwidth]{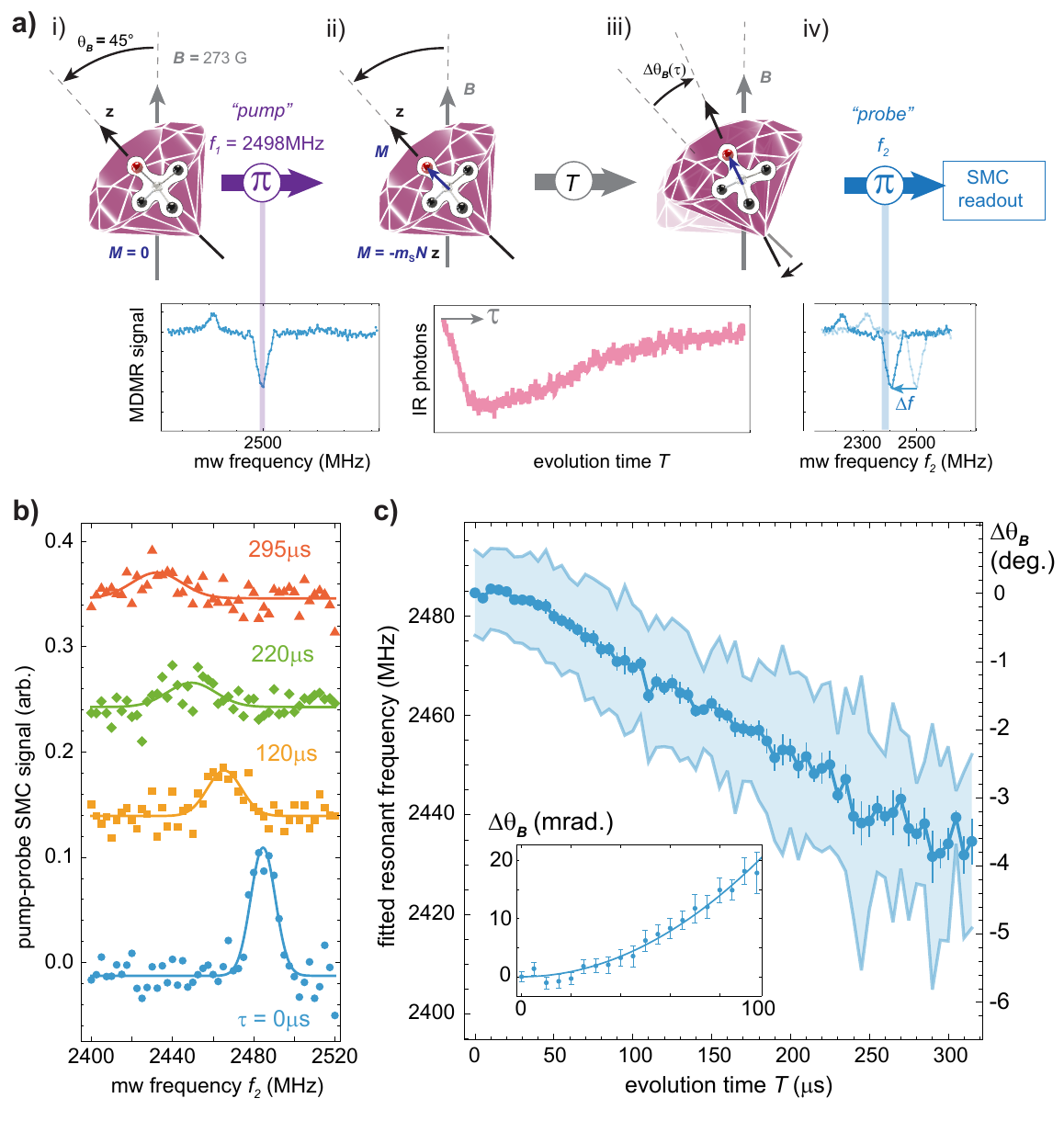}
	\caption{\textbf{Pump-probe measurement of angular displacement.} a) Schematic of experimental sequence and pertinent angles. A resonant `pump' microwave $\pi$-pulse initiates reorientation of the diamond particle. After a time $t_d$, a second `probe' $\pi$ pulse with varied frequency is applied: when resonant with the NV two-level splitting at the new diamond angle, the probe pulse arrests the spin-torque induced reorientation by repopulating the non-magnetic $m_S = 0$ state and drives the MDMR contrast towards unity. b) SMC contrast vs probe microwave frequency $f_2$ for different $t_d$ times, exhibiting a marked frequency shift in addition to amplitude suppression resulting from $T_1$-induced depolarisation. c) extracted contrast maximum position (points) and widths (shaded lines) vs evolution time $t_d$. Inset: the initial $100\,\upmu$s of evolution is given by $\theta(T) \approx 2\tau_\text{sp}/I t_d^2$, where $I$ is the moment of inertia, allowing the spin-induced torque $\tau_\text{sp}$ to be measured. Error bars in (c) extracted from standard error in fitted gaussian means.}
	\label{fig:fig4}
\end{figure}

{\bf Pump-probe measurement of angular displacement and spin torque.} The reorientation of the diamond that leads to an MDMR signal also results in an energy shift due to the angular dependence of the NV Zeeman shift. We measured the frequency perturbation generated from the mechanical action imparted by a $\pi$-pulse using a `pump-probe' arrangement as depicted in Fig. \ref{fig:fig4}(a). Here, a resonant `pump' $\pi$-pulse at $f_1 = 2498\,$\,MHz is first applied to initiate reorientation of the particle. A variable time delay $t_d$ is then inserted, allowing the particle to rotate by some angle $\theta_B$ towards the applied magnetic field, followed by application of another `probe' microwave $\pi$-pulse at frequency $f_2$. In the absence of any variation in Zeeman shift between the pump and probe pulse -- or zero delay time -- the NV ensemble will be returned to the $m_S = 0$ state by the second pulse and the spin-torque mediated reorientation of the particle will be arrested, and the MDMR NIR signal will approach unity. For longer times, the reorientation-induced frequency shift is measured by varying the frequency $f_2$ of the probe pulse.

Figure \ref{fig:fig4}(b) shows the pump-probe MDMR contrast signal as a function of $f_2$ for various delay times $t_d$, revealing a shift of more than 50\,MHz over 300\,$\upmu$s. Also evident is a significant drop in amplitude, a result primarily of $T_1$-induced depolarisation reducing the number of spins flipped by the probe pulse. Nevertheless, tracking the MDMR signal peak as a function of $t_d$ enables us to determine the angular displacement subtended by the particle during the free evolution time, shown in Figure \ref{fig:fig4}(c). By fitting the complete MDMR spectrum to a model with free parameters for the orientation of each NV axis to the applied magnetic field (see Supplementary Information), we deduce that the NV axis, originally $\sim 45^\circ$ to the 273\,G magnetic field, rotates by about $4^\circ$ or 70\,mrad in 300\,$\upmu$s. Focusing on the initial 100\,$\upmu$s of evolution (Fig. \ref{fig:fig4}(c, inset), where the angular displacement vs time is quadratic, simple kinematic arguments (See Supplementary Information) yield a torque of $\approx 6\times 10^{-17}$\,N\,m, implying we flip $1\times 10^8$ spins with the $\pi$-pulse.  

{\bf Discussion.}
The realisation of SMC and thus mechanical readout of pulsed spin manipulation in this work offers a vast array of established spin measurement and control protocols to be deployed in levitating spin systems and initial steps towards transducing coherent quantum dynamics into mechanical motion. Our results extend optomechanical detection of levitating spins into a new regime where the quantum state of a spin ensemble can be read-out nondestructively -- the NIR laser does not affect the particle motion or spin state -- and with vastly improved measurement contrast over traditional techniques. Only room-temperature ensemble readout techniques using mm-scale diamonds in resonant microwave cavities~\cite{eisenach_cavity-enhanced_2021} or laser-threshold readout~\cite{hahl_magnetic-field-dependent_2022, lindner_dual-media_2024} yield similar contrast values. SMC works best with diamond samples containing a high NV density, making it potentially beneficial for sensing as it is not affected by many of the issues that vex PL detection, such as background fluorescence, fluorescence overlap between NV charge states and limited photon collection efficiency~\cite{barry_sensitivity_2020}. Indeed, the ability to rotate the diamond quickly~\cite{jin_quantum_2024, wood_magnetic_2017} could enable improved magnetometry via rotational upconversion of DC fields~\cite{wood_$T_2$-limited_2018, wood_dc_2022}. Further, photon collection can be spectrally narrowed to the exact wavelength of the detection laser -- which can be essentially any wavelength provided it does not interact with the NV spin -- to eliminate background counts.

The results presented in this work also show that SMC serves as an interesting tool to better understand and ultimately control the host mechanical oscillator. A natural extension to consider is the regime where the NV centres remain spin-coherent over the duration of particle reorientation. In a simplified picture, a coherent superposition of spins should in principle drive the host particle into a coherent macroscopic superposition state of orientations. However, such an experiment requires the mechanical state decoherence time to comparable to the spin coherence time, which in turn requires high vacuum and even shielding from black-body radiation. Further, in the $N$-spin uncorrelated ensemble system we focus on in this work, a $\pi/2$-pulse applied to the spins creates the tensor product state, \emph{i.e.} $|\psi\rangle = 1/\sqrt{2}(|0\rangle + |1\rangle)^{\otimes N}$. To evolve the particle into a superposition of orientations, the spins should ideally be in an $N$-spin GHZ state, $|0,0,0...0\rangle + |1,1,1...1\rangle$ (See Supplementary Information), a tiny fraction of the tensor product state and a far harder experimental prospect to realise. 

Nevertheless, if the NV spin ensemble were put into a coherent superposition state that persisted over the time required to realise measurable angular displacement of the diamond, on the order of $100\,\upmu$s, then the effects of coherent transverse spin dynamics on the particle motion may begin to be apparent. In the magnetic fields used in these experiments, the transverse spin component of the NV ensemble precesses at GHz frequencies and so does not significantly couple to the particle motion. However, by applying large magnetic fields that approach the ground-state level anti-crossing (GSLAC)~\cite{perdriat_angle_2022}, lower precession frequencies on the order of the trap secular frequencies may be possible. Further, a range of sophisticated dynamical decoupling pulse sequences exist that extend spin coherence times from a few $\upmu$s by ten or hundredfold, especially in NV-dense, highly disordered diamond samples~\cite{zhou_quantum_2020, zhou_robust_2023} such as that used here. Feed-forward of the microwave detuning to retain spin resonance~\cite{wood_quantum_2021} can be employed to realise coherent control in periodic time-varying magnetic environments. On the materials engineering front, commercially available CVD diamond substrates with coherence times approaching $100$-$200\,\upmu$s and NV densities of $\sim300\,$ppb could be milled into smaller micro or nanodiamonds~\cite{wood_long_2022} than the $10\,\upmu$m samples used here, such that lower defect concentrations yield comparable reorientation angles without significant degradation of spin properties as observed in nanodiamonds.     

In conclusion, we have demonstrated spin-mechanical conversion as a method for readout of a levitated quantum spin ensemble. Our technique allows for mechanical detection of Rabi oscillations, spin-echo interferometry and $T_1$ relaxation, with readout contrast better than that of a single NV centre in an ultrapure diamond and more than 50 times better than standard PL readout under the same conditions. These results are a powerful enabling technology for levitated spin systems, and open up new opportunities for precision sensing and studies of macroscopic quantum phenomena.

{\bf Methods.} Our experiment consists of a Paul trap formed from a 25\,$\upmu$m tungsten wire curved into a horseshoe shape. The Paul trap is mounted on a three-axis piezo nanopositioner to optimise alignment with the illumination beam. An AC potential of $1.5$-$2\,$kV at $2$-$3$\,kHz is applied to the Paul trap electrode while a bias-tee allows microwave current (2.2-4\,GHz) to flow through the loop shape. Microdiamond particles with typical diameters of $10$-$12\,\upmu$m containing 1-2\,ppm of NV centres (Adamas Nano) are trapped at the neck region of the trap where microwave coupling is strongest, with typical librational trapping frequencies of 100-500\,Hz. All our experiments are conducted at atmospheric pressure, but inside a vacuum chamber with passive dehumidification to prolong trap lifetime. Green light (561\,nm) and 780\,nm NIR light is focused onto the trapped particle using a 0.6-NA microscope objective with an 11-mm working distance. Detection of both scattered 780\,nm light and NV PL (650-800\,nm) is performed with a single-photon counting module (SPCM), and a 750\,nm long-pass filter is used to isolate PL from scattered NIR light. The PL and scattered photon flux is attenuated by a factor of $10^4$ with a neutral density filter to prevent damage to the SPCM. 

Microdiamonds are loaded into the trap by placing a charged injector wire coated in microdiamond powder near the electrodes. Consequently, samples loaded into the trap exhibit a range of shapes, sizes and a significant spread of NV densities. For each particle loaded into the trap, the AC voltage and frequency are adjusted to ensure the particle is in a stable librational state of motion rather than rotating~\cite{perdriat_angular_2023} and is resistant to sustained green illumination. Many trapped particles exposed to even modest green laser intensities quickly become unstable and exit the trap, the reasons are not fully understood but are assumed to be due to surface charge stability. Trapping voltages and frequencies thus vary for each of the four samples we present data for in this work, and are documented in the Supplementary Information.

{\bf Theory.} We consider a microdiamond trapped in an electric Paul trap hosting an ensemble of NV centres. In what follows, we ignore interactions between NV centres and assume a purely classical description of the particle motional dynamics. The simplified Hamiltonian describing an NV spin in the presence of a magnetic field $\mathbf{B}$ is given by
\begin{equation}
H/\hbar  = D_\text{zfs} S_z^2 + \gamma_e\mathbf{B}\cdot\mathbf{S},
\label{eq:eq1}
\end{equation}
with $D_\text{zfs}/2\pi = 2.87\,$GHz, $\gamma_e/2\pi = 2.8\,$MHz/G and $\mathbf{S} = (S_x, S_y, S_z)$ the vector of spin-1 Pauli matrices $S_i$. We assume the orientation of the diamond is trapped in an approximately harmonic potential with trapping frequency $\omega$ and librational degree of freedom $\theta$ obeying    
\begin{equation}
I \ddot{\theta}+I\gamma_g\dot{\theta}+I\omega^2\theta = \tau(t).
\label{eq:eq2}
\end{equation}
Here, $I$ is the particle moment of inertia, $\gamma_g$ a gas damping coefficient and $\tau(t)$ a time-dependent torque containing a stochastic contribution $\tilde{\tau}(t)$ due to brownian motion and a spin-mechanical torque
\begin{equation}
\tau_\text{spm}(t) = -\text{Tr}\left(\rho(t)\frac{\partial H}{\partial \theta}\right)
\label{eq:eq3}
\end{equation}
originating from the NV spin polarisation. The observable particle dynamics result from an interplay between the harmonic trapping potential (including particle micromotion), spin depolarisation and librational motion in the overdamped regime. Nevertheless, we observe that spin-induced reorientation of the particle leads to significant particle rotation with sufficient reproducibility to serve as a high fidelity readout of the NV ensemble quantum state. Calculations of expected spin torque induced rotations are provided in the Supplementary Information.

\section*{Acknowledgments}
This work was supported by the Australian Research Council (DP240100942). 

\section*{Author contributions}
A. A. W. conceived and led the work with theory support from A. M. M. The experiment was built by A. A. W., R. M. G. and T. X. with contributions from T. D. and G. H. Data was collected and analysed by A. A. W. and T. X. Computation of the MDMR spectra for angle tracking measurements was conducted by D. S. R. Theory for spin-mechanical dynamics was developed by F. H. C., A. M. M. and T. D, and A. A. W wrote the manuscript with contributions from all authors.

\section*{Competing interests}
The authors declare no competing interests.

\section*{Data availability}
All raw data and analysis code is available from the corresponding author (A. A. W.) upon reasonable request. 

\newpage
\widetext
\begin{center}
\textbf{\large Measurement of a quantum system using spin-mechanical conversion - SUPPLEMENTARY INFORMATION}
\end{center}

\setcounter{equation}{0}
\setcounter{figure}{0}
\setcounter{table}{0}
\setcounter{page}{1}
\makeatletter

\renewcommand{\thesection}{S\arabic{section}}
\renewcommand{\thefigure}{S\arabic{figure}}
\renewcommand*{\citenumfont}[1]{S#1}
\renewcommand*{\bibnumfmt}[1]{[S#1]}

\section{Experimental setup and details}

\begin{figure*}
	\centering
		\includegraphics[width = \textwidth]{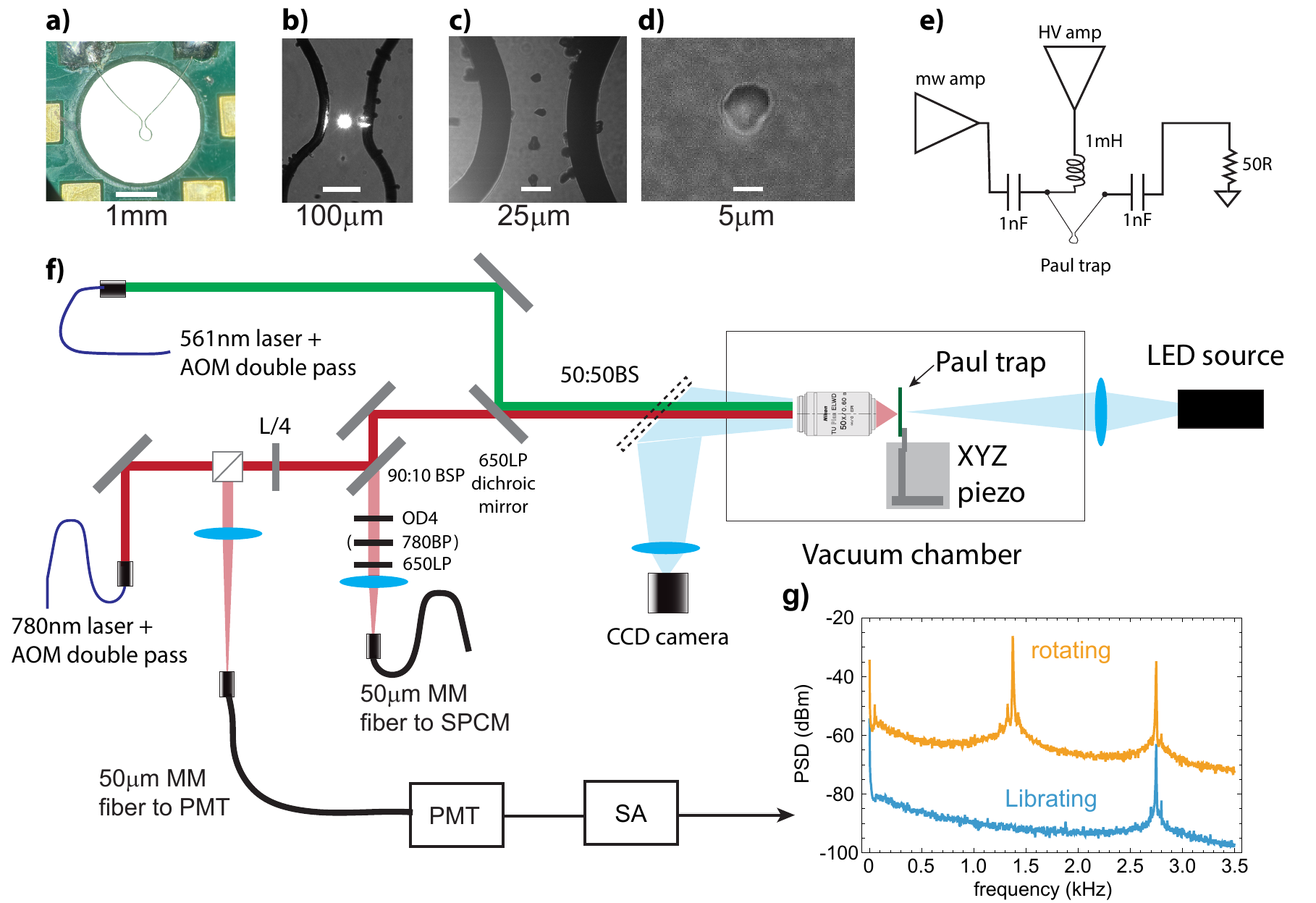}
	\caption{Schematic of the experimental setup. a) photograph of the Paul trap wire soldered to a printed circuit board, and b-d) images of trapped microdiamonds at different scales. In a), the scattered green light from a single particle is shown, in c) multiple particles are trapped, and in d) a high-magnification image of a single microdiamond is shown. Particles are illuminated with light flashed synchronously with the trap AC drive frequency to eliminate micromotion. e) Circuit to simultaneously apply high-voltage signal to Paul trap electrode and transmit microwaves. f) optical setup and beam paths. g) Power spectral density of particle motion measured via PMT and spectrum analyser, for a rotating particle (orange) and a librating particle (blue), traces vertically offset for clarity.}
	\label{fig:figs1}
\end{figure*}

The experimental setup consists of a microdiamond trapped in a home-made Paul trap, Fig.~\ref{fig:figs1}(a-d). Microdiamond samples are sourced from Adam\'as Nanotechnology and are 8-12\,$\upmu$m in diameter and specified to have NV densities of between 1-2\,ppm. We typically observe that between 10-20\% of trapped particles exhibit markedly lower fluorescence and consequently no observable MDMR signal.

{\bf Paul trap.} The Paul trap itself is fabricated by hand using 25\,$\upmu$m gold-plated tungsten wire first deformed into a V-shape, the ring bottom and bottleneck trapping regions are created by pinching the wire encompassing a length of 50\,$\upmu$m wire. The Paul trap is soldered onto a custom PCB that provides mechanical stability and interfaces to the HV drive signal and microwaves. The complete assembly and circuit diagram is depicted in Fig.~\ref{fig:figs1}(a, e). The low-frequency HV signal (1-3\,kHz, 600-2000\,V) is sourced from the LF output of a vector signal generator (Rohde and Schwartz SMBV100A) and amplified by a Matsusada AMT-10-10 HV amplifier, which connects to the Paul trap via a 1\,mH high-voltage inductor on the PCB. Microwaves are combined with the AC drive signal using two 1\,nF 3000\,V capacitors. Thus, the HV low frequency Paul trap sees an open circuit, while the microwaves pass through the wire to ground via a 30\,dB 50\,$\Omega$ attenuator. The Paul trap is mounted on a three-axis piezoelectric translation stage (Xeryon, XLS-3-40-HV) capable of 25\,mm of travel in all directions with 5\,nm encoder resolution. The entire assembly is housed within a large vacuum chamber. Though no measurements were conducted below atmospheric pressure in this work, the chamber serves as essential protection against air currents which easily de-trap the diamonds.

{\bf Loading.} Microdiamonds are loaded into the Paul trap using the `injection' method described in Ref. \cite{sdelord_spin-mechanics_2019, sperdriat_angular_2023}. A PVC tube is charged via triboelectric agitation with a woolen garment, and an insulated probe tipped with 2\,cm of 100\,$\upmu$m tungsten wire is dragged across the charged PVC tube. Electrons from the wire are donated to the PVC, giving the metal an overall negative charge. The tungsten tip is then dipped into microdiamond powder, which readily adheres to the charged probe. Loading is then performed by positioning the microdiamond-coated tungsten tip near the Paul trap, which is usually operated at 2\,kV 3\,kHz during loading. When the probe approaches the trap, many microdiamonds are attracted to the bottleneck (see eg, Fig.~\ref{fig:figs1}(c)), ring trapping region and in most cases, the electrodes themselves. We select out one diamond by stimulated detrapping with air currents and fast agitation of the piezo stage: in the latter case, the ultrasonic piezo drive signal is very effective at removing diamonds adhered to the electrodes and loading them into the trapping region.

{\bf Optics.} A schematic of the optical paths in the experiment is depicted in Fig. \ref{fig:figs1}(e). A fixed microscope objective (Nikon TU Plan ELWD 0.6NA) focuses both green and NIR light onto the trapped diamond and simultaneously collects scattered NIR light and NV photoluminescence. A white light source filtered with blue transmission filters illuminates the Paul trap from behind, and a flipper mirror system allows the objective lens to form an image of the trapped particle. The 561\,nm laser is a diode-pumped solid state laser from CNI, and is gated with a double-passed AOM system before coupling into a single-mode, polarisation maintaining fiber. Green light is directed onto the back aperture of the microscope objective using a 638\,nm long pass dichroic mirror. The 780\,nm laser is a MOGLabs CatEye ECDL with $\sim 0.1\,$MHz linewidth that is also gated using a double-passed AOM. After coupling into a single-mode, polarisation maintaining fiber, the 780\,nm light passes through a PBS cube and quarter-wave plate (forming an optical valve) and combined with the green light on the dichroic mirror. NV PL (600 - 800\,nm) and scattered 780\,nm light from the particle are both collected by the microscope objective. A 90:10 beamsplitter leaks some of this light into a collection setup, which features an OD4 filter to attenuate NV PL and IR photon flux, a switchable 780\,nm bandpass filter to select only scattered IR light and a 650\,LP filter to prevent green light reflecting into the detectors when configured for PL detection. A 100\,mm focal length achromatic lens couples the light into a 50\,$\upmu$m core multi-mode fiber and sends it to a single photon counting module (Excelitas SPCM-AQRH-14-FC). Downstream of the SPCM detection, IR light that is polarisation rotated by the quarter-wave plate exits via the reflection port of the PBS cube, where it is coupled into a 50\,$\upmu$m core fiber and directed into a photomultiplier tube (Hamamatsu R928). The PMT signal is connected to a spectrum analyser (Moku:Lab) and a power spectral density of the particle's motion and rotation is continuously acquired.

{\bf Stabilisation.} After a microdiamond is loaded into the trap, we measure the particle motion using scattered 780\,nm light on the PMT. Every diamond has a different shape and charge-to-mass ratio, and generally traps in different spatial positions within the Paul trap bottleneck. Therefore, a loaded diamond may exhibit considerable micromotion (translation in any direction at the trap drive frequency) and/or exhibit rotation at half the trap frequency~\cite{sperdriat_rotational_2024}. Appearance of a feature at half the Paul trap drive frequency is an easily identifiable hallmark of rotation, as shown in Fig. \ref{fig:figs1}(g). To stabilise the particle, we combine slow reduction of the drive voltage with increases to the trap frequency to eliminate rotation (see Ref. \cite{sperdriat_rotational_2024}). Occasionally, sudden changes of the drive voltage from high to low at fixed frequency can also be used to transition the particle from the rotating to librating regime. Each trapped microdiamond examined in this work is therefore held at slightly different trap drive frequencies based on what parameters stabilised the librational dynamics after loading, as shown in Table \ref{tab:1}.

\begin{table}
\caption{\label{tab:1} Paul trap parameters used in experiments.}

\begin{tabular}{ccc}
Figure in main text & $f_\text{AC}$ (Hz) & $V_\text{AC}$ (V$_\text{pp}$)\\
\hline
\hline
Figure 1 data (all) & 2375 & 1840\\
Figure 2 (a) & 2790 & 1950\\
Figure 2 (b) & 2375 & 1840\\
Figure 3 (all) & 2895 & 1632\\
Figure 4 (all) & 2790 & 1950\\
\hline
\hline
\end{tabular}
\end{table} 
\section{Theoretical description}
\begin{figure*}
	\centering
		\includegraphics[width = \textwidth]{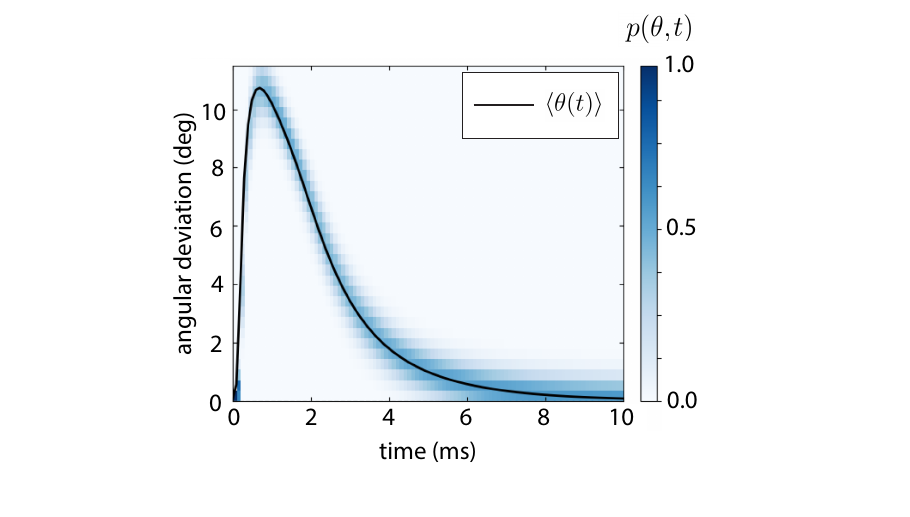}
	\caption{Simulated marginal angular probability density function (blue) of a levitated microdiamond when coupled to NVs and allowed to evolve freely, modelled with Brownian noise, with the ensemble average overlaid (black). Simulations conducted for an ellipsoidal ($\frac{a}{b}=0.5$) diamond of nominal radius $10\mu$m, with $10^8$ NVs, decoherence time of of $T_1=7$\,ms in a harmonic trap of frequency $2300$rad/s, and atmospheric pressure resulting in a gas damping factor of $\gamma_g=6280$\,rad/s}
	\label{fig:figst}
\end{figure*}

The theory introduced in the main text (Eqs. 1-3) reproduces the observed experimental results, specifically the characteristic particle deviation observed in Fig. 1(d). In the experiment, a pulse of green light is used to polarise the NV into the $m_S = 0$ state, followed by a microwave $\pi$-pulse, flipping the spin into a magnetic state. The levitated diamond will then rotate towards alignment or anti-alignment of the NV axis with the external field. Each NV in the participating orientation class will apply a torque of $\tau(t) = |\mu_\text{NV}|B\sin(\phi-\theta(t))$, where $\mu_\text{NV}$ is the NV magnetic moment, approximately equal to the gyromagnetic ratio for an electron, and $\phi$ is the angle between the magnetic field and the NV axis equilibrium alignment. $T_1$ decoherence will cause the torque to decay as a thermal mixture of populations is established, and can be phenomenologically modelled by incorporating a decay term, $\exp(-\frac{t}{T_1})$. Thus, the resulting equation of motion for the particle's harmonically trapped librational mode with degree of freedom $\theta$ is
\begin{align}
    I\ddot\theta + I\gamma_g\dot\theta+I\omega^2\theta = Ne^{-\frac{t}{T_1}}|\mu_{NV}|B\sin(\phi-\theta(t)) + \tilde{\tau}(t)
		\label{eq:eom}
\end{align}
where $\omega$ the harmonic trapping frequency of the relevant librational mode of the trap, $\gamma_g$ a gas damping rate, $I$ the moment of inertia, and $N$ is the number of NV spins flipped by the microwave pulse. As the experiments were not performed at vacuum pressure, there is a Brownian noise effect on the levitated diamond due to random collisions with air particles. This can be modelled as an additional torque \cite{sperdriat_angular_2023}, $\tilde{\tau}(t)$, with properties
\begin{align*}
    \langle\tilde{\tau}(t)\rangle = 0\,\,\,\,\,\,\,\,\ \langle \tilde{\tau}(t)\tilde{\tau}(t')\rangle &=2\gamma I k_B T \delta(t-t'),
\end{align*}
where $T = 300\,$K is the ambient temperature, $k_B$ is the Boltzmann constant and $\delta(t-t')$ the Dirac delta function. Inclusion of a noise term forbids a deterministic solution to the equation of motion. Instead, the Fokker-Planck equation can be used to solve for the probability distribution function (PDF) $p(\theta, \dot\theta,t)$ for the angular position and velocity. For this system, the Fokker-Planck equation is
\begin{align}
    \partial_t p(\theta, \dot\theta,t) = -\partial_\theta \left(\dot\theta \, p(\theta,\dot\theta,t)\right)
    +\partial_{\dot\theta} \left[\left(\gamma\dot\theta+\omega^2\theta-\frac{\tau(\theta,t)}{I}\right)\, p(\theta, \dot\theta,t)\right]
    +\partial^2_{\dot\theta} \left(\frac{ \gamma}{\beta}\, p(\theta, \dot\theta,t)\right),
\end{align}
with $\beta = \frac{1}{k_B T}$. Simulations of the particle angular deviation were done for an ellipsoidal diamond of nominal radius $10\mu$m, with $10^8$ NVs and $T_1=7$\,ms in a harmonic trap of frequency $\omega = 2300$rad/s and a gas damping rate of $\gamma_g=6280$\,rad/s. Fig.~\ref{fig:figst} shows the Fokker-Planck solution for the probability density of the alignment angle over time, integrated over the angular velocity ($\dot\theta(t)$). Overlaid on this figure is the ensemble average angular position $\langle\theta(t)\rangle$, referred to as the first moment, defined as
\begin{align}
    \langle\theta(t)\rangle = \int_{-\infty}^{\infty}\theta(t)\cdotp p(\theta,\dot\theta, t) d\theta.
\end{align}
The theoretical simulations closely resemble what is observed in experiments. Estimation of the trapping parameters at atmospheric pressure is complicated by the inability to resolve secular motion or libration due to significant gas damping, although the trapping parameters we use here are reasonable for the trap we use~\cite{sdelord_spin-cooling_2020}.

\section{Micromotion}
Particles trapped in the Paul trap exhibit motion at the trap AC drive frequency, called \emph{micromotion}, which arises because the particle is never trapped at the exact quadrupole zero of the electric field. Gravity and stray electric fields displace the particle and result in translational micromotion. Despite the name, the micromotion can be significant in experiments and well in excess of the slower secular motion, leading to visible blurring of the particle. No active reduction of micromotion was attempted in our experiments, despite it being omnipresent and by far the dominant particle motion. Secular motion or angular dynamics at the harmonic trapping frequencies is difficult to observe due to the strong damping regime at atmospheric pressures. 

\begin{figure*}
	\centering
		\includegraphics[width = \textwidth]{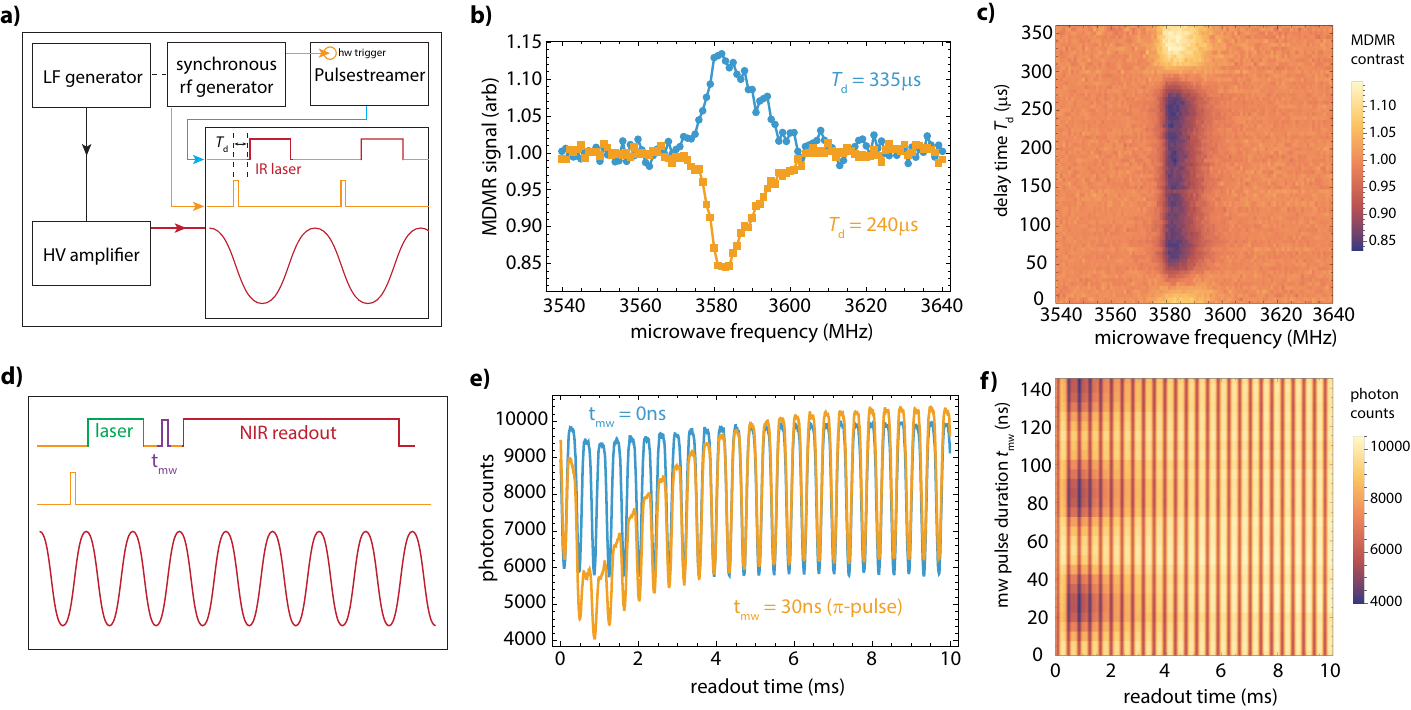}
	\caption{Probing the effects of micromotion on SMC readout. a) experimental schematic of the implementation of synchronous pulsing with AC Paul trap drive. NIR light is pulsed synchronously with the AC drive, and a delay time $T_d$ allows different phases of the micromotion to be selected. b) MDMR signal for two different delay times, exhibiting inverted contrast, and c) variation of MDMR signal as a function of delay time. d) Conducting SMC readout experiments synchronous with the micromotion using a similar implementation as in a), except that micromotion is sampled during the 10\,ms NIR readout in addition to the spin-torque signal in e), f). While micromotion is clearly evident, the spin-torque signal can nevertheless be revealed. }
	\label{fig:figs2}
\end{figure*}

While measurements conducted in the main text (notably Fig 3) show evidence of micromotion induced by both the green laser initialisation and spin torque, intrinsic micromotion from the trap drive is absent due to the lack of any synchronisation between experiments and trap drive frequency. As such, micromotion is largely averaged out after several experimental repetitions and does not influence our measurements. We examined micromotion's effect by synchronising experiments to the trap AC frequency. A synchronous signal is used to hardware-trigger the experimental pulse generator (Swabian Instruments PulseStreamer 8/2) so that laser and microwave pulses are applied at well-defined phases of the trap drive, meaning that micromotion is no longer averaged out. A diagram showing how synchronisation is implemented is depicted in Fig. \ref{fig:figs2}(a). First, we simply pulse the 780\,nm laser for 10\,$\upmu$s a fixed delay time $T_\text{d}$ after the synchronisation trigger while green light and microwaves are applied continuously. In this manner, the NIR light stroboscopically measures the MDMR signal at a specific phase of micromotion, which in this case is at the trap drive frequency of 2778\,Hz. Notably, the MDMR spectra can be quite different from the overall averaged feature we typically see in unsynchronised measurements, exhibiting positive and negative contrast at different delay times as shown in Fig. \ref{fig:figs2}(b,c). 

Next, we synchronise time-resolved measurement of SMC Rabi oscillations to the trap micromotion (Fig. \ref{fig:figs2}(c)). Following a 500\,$\upmu$s green laser pulse (which initialises the NVs over several micromotion periods, $f = 2564\,$Hz) a variable duration microwave pulse is applied before NIR readout of the particle motion is initiated. Strong modulation of the reflected NIR light at the trap frequency is evident in the measured time traces (Fig. \ref{fig:figs2}(d,e)), though the effects of a microwave $\pi$-pulse still result in a significant reduction in signal comparable to what we see in averaged, unsynchronised experiments (Fig 1 of the main text). These measurements confirm the presence of significant micromotion, but also show that it does not affect the spin-torque measurements described in the main text. One significant effect of micromotion is that in unsynchronised measurements it demands multiple experimental averages to be suppressed, which makes single-shot readout (SSR) of spins via SMC difficult, despite exhibiting photon count rates ($>1000$ photons per bin when accounting for the neutral density filters used to attenuate photon flux) and spin readout contrast ($70$\% as reported in the main text) that should make SSR possible. However, synchronisation to the trap drive frequency and limiting collection to a fixed point of the modulated readout signal should serve to account for micromotion, making SSR feasible. 

\section{Optical perturbations}
In the main text, we observed in Fig 3 evidence that pulsing the green laser on resulted in a perturbation, visible as micromotion at the AC drive frequency. In other experiments we saw clear evidence that the green light results in a position and sometimes angle displacement of the diamond. Given the central tenet of SMC is that green light can be pulsed intermittently to optically prepare the NV spin, it is thus essential to examine the perturbative effect of the green light, and to some extent as well, the NIR readout light. While the NIR light can be reduced to well below 1\,$\upmu$W for readout, higher green powers (typically more than 50\,$\upmu$W) are needed in order to spin-polarise the NV ensemble, particularly when the laser is pulsed only briefly for Rabi measurements. The perturbative effect of the green laser thus dominates over the NIR, and stems primarily from radiation pressure~\cite{sdelord_electron_2017}.   

Some microdiamonds are particularly perturbed by application of green light, visibly moving to a new position when the light is applied at powers of order 50\,$\upmu$W. These diamonds are typically difficult to work with, unstable and consequently discarded. Other diamonds exhibit the converse, with imperceptible changes in the scattered IR light when green light is applied. While larger diamonds (the specified size range is 8-12\,$\upmu$m) are generally less perturbed by the green light than smaller particles, we suspect morphology has the most significant bearing on the observed perturbations. Morphology is difficult to quantify, as trapped particles are generally held in one orientation in the trap and reorientation is not possible to observe the particle aspect ratio, for instance particles that appear large in one direction may be flat and flake-like, and thus strongly perturbed by the laser. We present here some extended data on how green light perturbs trapped particles, noting that data presented in the main text was generally extracted from very stably trapped particles that exhibited minimum perturbation from the applied green light.

{\bf SMC vs. CW MDMR}

\begin{figure*}
	\centering
		\includegraphics[width = \textwidth]{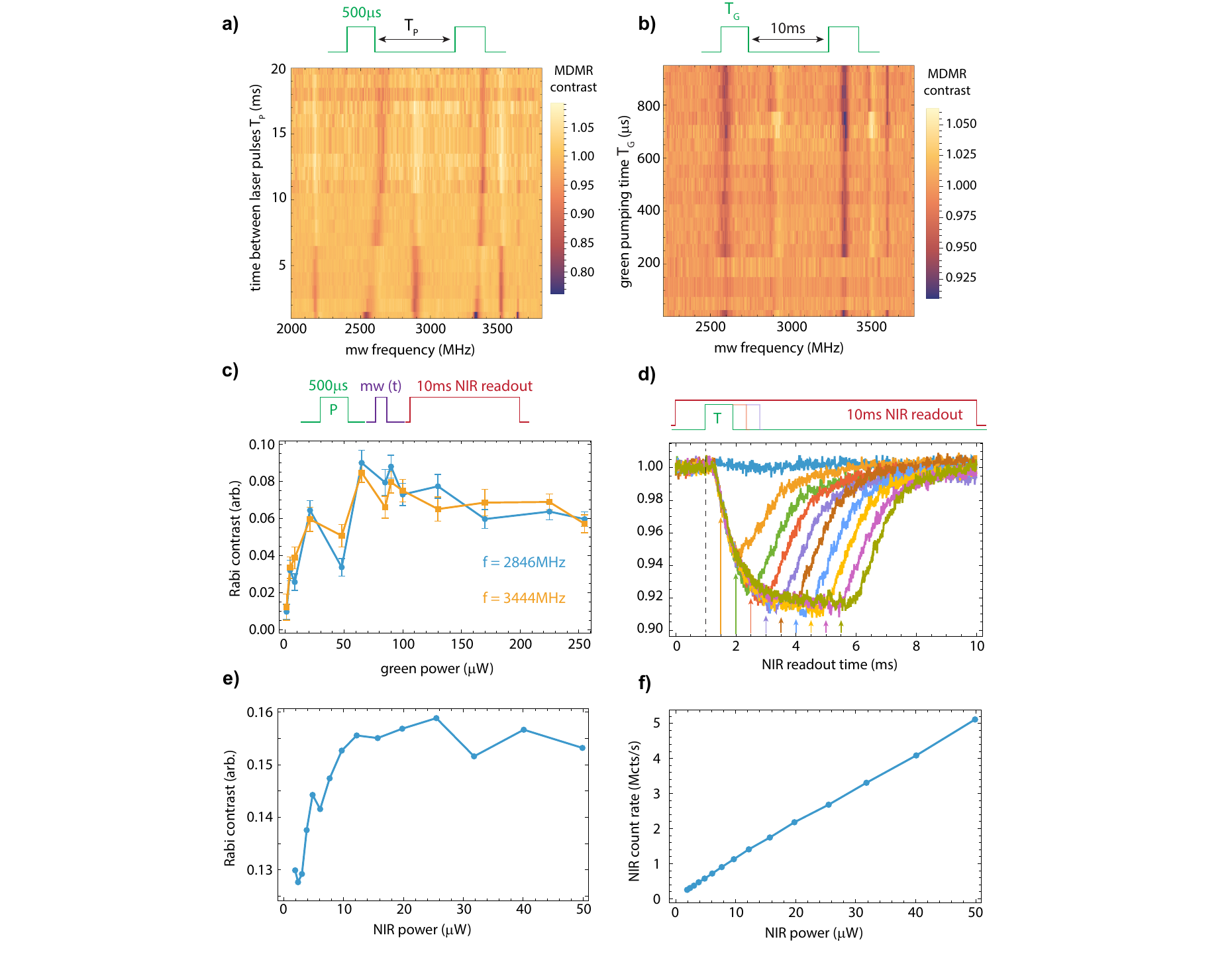}
	\caption{Evaluating the effect of green light, a) cw-MDMR vs frequency signal as a function of time between green laser pump pulses (500\,$\upmu$s, 60\,$\upmu$W), exhibiting variation in both contrast and frequency due to motional perturbation caused by the green laser. b) Varying the duration of the green laser pulse (constant power of 60\,$\upmu$W) for fixed interpulse time (10\,ms) has a less significant effect on the particle motion and hence MDMR signal. c) SMC Rabi oscillation amplitude as a function of green laser power for two different orientation classes and frequencies. Laser powers above 50\,$\upmu$W are required for sufficient spin polarisation to effect an MDMR signal. d) Pulsing a variable duration green laser pulse in the midst of NIR recording allows the pertubation of the green laser to be observed directly. Vertical arrows denote when corresponding green pulse finishes: note that the particle continues to move after the light is extinguished. For c) and d), a 532\,nm laser was used due to failure of the 561\,nm source. e) SMC Rabi contrast as a function of NIR power, and f) steady-state NIR count rate as a function of NIR power.}
	\label{fig:figs3}
\end{figure*}

In MDMR, continuous application of the green light leads the particle to adopt a steady state orientation that changes quasi-statically when resonant microwaves are applied. For particles notably perturbed by green light, we observe that the particle orientation under cw green illumination leads to a different MDMR spectrum than if the green light is pulsed at the duty cycles used in SMC and Rabi oscillations, stemming from the different time-averaged force applied by the green laser. Therefore, the resonant frequency identified in a cw-MDMR measurement may not be the correct frequency when the green laser is pulsed to execute a Rabi measurement, or the MDMR contrast can vary significantly, sometimes going from negative to positive. This effect is illustrated in Fig. \ref{fig:figs3}(a,b), where we apply continuous microwaves but pulse the green laser (50\,$\upmu$W) for different times and different rates. While varying the laser pulse length at a fixed repetition rate of 100\,Hz has only a modest influence on MDMR contrast, varying the repetition rate significantly alters the contrast and also perturbs the frequency of MDMR features.

Next, we directly assess the impact of varied green power on the measurement of mechanically-detected Rabi oscillations. For a fixed green pumping time of 500\,$\upmu$s, we vary the green pumping power and measure the amplitude of the Rabi oscillations, depicted in Fig. \ref{fig:figs3}(c). As expected, for low powers of a few $\upmu$W spin polarisation is incomplete within a 500\,$\upmu$s pulse, leading to low amplitude Rabi oscillations. Beyond about 100\,$\upmu$W, the Rabi contrast saturates, indicating saturated spin polarisation. 

Next, we examined the effect of the green laser on the particle motion directly, recording the scattered NIR light during application of a 50\,$\upmu$W green pulse, applied 1\,ms after recording begins in Fig. \ref{fig:figs3}(d). The transient particle motion imparted by the green pulse exhibits an interesting effect, in that the particle continues to move for 500\,$\upmu$s or more after the green laser is extinguished, in the same direction the laser perturbed it, before returning to the equilibrium position. This is in contrast to the experiments reported in the main text in Fig 4, where upon application of a microwave $\pi$-pulse the motion is arrested abruptly. Radiation pressure forces from the green laser should also terminate abruptly upon extinguishing the laser. Due to spin mixing between NV $m_S$ levels in the presence of a large, off-axis magnetic field, a pure $m_S = 0$ state exhibits a nonzero magnetisation, though it is at most $\sim 1.5$\% of that for a fully-polarised $m_S = \pm1$ state. Further experiments are needed to examine this effect, though it has at most a minor role in the current realisation of SMC. We also note here that due to failure of the 561\,nm laser source used in experiments, a 532\,nm green laser was used, although this substitution is expected to have no significant change on the experiment.   

We also studied the effects of varied NIR power.  In Fig. \ref{fig:figs3}(e), we plot the SMC Rabi contrast as a function of NIR power, which exhibits only a minor change across 50\,$\upmu$W of variation. We attribute the slightly lower contrast at powers below 10\,$\upmu$W, from 13\% to 16\%, to greater noise in the lower power data (all data was taken for fixed 3\,min averaging time) rather than NIR power induced mechanical effects which appeared to be absent. In Fig. \ref{fig:figs3}(f) we show the steady-state (i.e. continuous NIR illumination) count rate as a function of the same NIR power variation. Noting that count rates into the APD are attenuated by a factor $10^4$, the NIR powers required to read out the mechanical state of the particle can be significantly below what we use here. The choice to attenuate counts was purely a matter of convenience so that PL detection can be used. PL count rates from the green illumination of the diamond (which cannot be attenuated, since spin polarisation would suffer) are estimated to be $>20\times10^9$\,cts/s and would damage the APD detector. Thus, PL counts must be attenuated before entering the APD, and it proved simpler to use the same detection line for PL and NIR light.
 
\section{Measurement sensitivity} 

Using spin mechanical conversion as a means of reading the outcome of a quantum measurement on a group of spins offers an alternative to conventional fluorescence detection. We consider the case where green light is used to prepare a spin ensemble into an $m_S = 0$ state, followed by a sequence of microwave pulses designed to convert the relative phase acquired between start and end of a sequence into a population difference between $m_S = 0$ and $m_S =\pm1$ states, such as Ramsey or Hahn-echo type interferometry. In what follows, MDMR is assumed to act purely in a readout capacity, and the case where the field to be sensed induces a spin-torque itself has already been examined in Ref. \cite{sperdriat_angular_2023}. The enhanced readout contrast conferred by MDMR, in principle 100\%, is tempered by the necessity to wait for timescales on the order of several ms for the spin-torque to yield an appreciable particle deflection, and then to subsequently recover the initial orientation for another measurement. Beyond shot noise, which is the dominant noise source, we must consider how effects that result in a noisy angular deviation -- such as spin projection noise and thermal fluctuations of the diamond orientation -- map to the detected photon rate. 

\subsection{Photon shot noise}

The measurement we perform in the main text is one where spin polarisation (and thus the outcome of a quantum measurement on the ensemble of spins) is mapped to the intensity of scattered NIR light on a detector. Assuming the collected photons vary linearly between two values, $x_{|0\rangle}$ and $x_{|-1\rangle}$ corresponding to angular extremes for maximally polarised spin states, $\theta\in (\theta_0$, $\theta{-1})$, \emph{photon shot noise} will result in a minimum detectable angular variation. 

A Ramsey-type measurement on the spin ensemble will yield an SMC signal given by 

\begin{equation}
S = (x_{|0\rangle}-x_{|-1\rangle})\cos^2\left(\frac{\gamma_e \delta B t}{2}\right)
\label{eq:4}
\end{equation}
with $\delta B$ the small magnetic field to be sensed, $\gamma_e$ the electron gyromagnetic ratio and $t$ the free evolution time. Assuming operation at the mid-fringe point, the linearised Ramsey signal will be 
\begin{equation}
S = (x_{|0\rangle}-x_{|-1\rangle})\frac{\gamma_e \delta B t}{2}.
\label{eq:3}
\end{equation}
At the mid-fringe point, we detect $x_{|-1\rangle}+\Delta x/2$ photons, where $\Delta x =(x_{|0\rangle}-x_{|-1\rangle})$. Let us assume that $x_{|-1\rangle} \ll \Delta x/2$, \emph{i.e.} that the normalised readout contrast is very close to one, not an inconceivable prospect for SMC. Thus, the signal error will be $\delta S \approx \sqrt{\Delta x/2}$ and 
\begin{equation}
\frac{dS}{dB} = \frac{\gamma_e T \Delta x}{2}, \hspace{5mm} \delta B = \frac{2\delta S}{\Delta x \gamma T} = \sqrt{\frac{2}{\Delta x}}\frac{1}{\gamma_e T}.
\label{eq:2}
\end{equation}
The photon shot noise will depend on the number of measurements we average for. Assuming a Ramsey-type measurement for $T_2^\ast \ll t_\theta \ll t_d$, with $t_\theta$ the time over which the diamond rotates under the spin torque and $t_d$ the dead time required to pump and reset the initial state of the diamond's motion, the number of measurements possible per second is $1/t_d$, and the photon shot noise-limited sensitivity is 
\begin{equation}
\delta B = \sqrt{\frac{2 t_d}{\Delta x}}\frac{1}{\gamma_e T}.
\label{eq:1}
\end{equation}
The steady-state NIR photon flux with a 50\,$\upmu$W impingent laser from a levitated diamond can be as high as $1\times10^7$ photons/s, \emph{after} an ND$=4$ attenuator ($\sim 10^4$ reduction). In a single measurement we feasibly collect upwards of 10000 photons in a 1\,ms-wide bin that contains the spin contrast information. With $T = T_2^\ast = 100\,$ns, $t_d = 10\,$ms, $\Delta x/2 = 5000$ the shot-noise limited sensitivity is poor, only $\delta B = 500\,$nT/$\sqrt{\text{Hz}}$. Accounting for the ND filter, the shot noise limit improves to 5\,nT/$\sqrt{\text{Hz}}$.

A comprehensive strategy to optimise SMC for improved magnetometric sensitivity is beyond the scope of this paper, involving detailed consideration of trapping potential, diamond material (chiefly size and NV concentration, as well as spin coherence properties) and so on. The principal limiting factors are the low measurement duty cycle due to the long mechanical relaxation times and the intrinsically poor spin coherence times of the NV centres in the sample. An increase to $T_2^\ast \sim 1\,\upmu$s (or potentially a $T_2\sim 10\,\upmu$s-limited Hahn-echo sequence) is feasible, and a librational frequency on the order of $1-10\,$kHz ($t_d \sim 100\,\upmu$s) may be possible in a needle-type Paul trap geometry~\cite{sskakunenko_strong_2025}, the corresponding sensitivity increases to 50\,pT/$\sqrt{\text{Hz}}$. Improving the photon collection efficiency is eminently feasible with a dedicated optical setup, and generating more photons via higher laser power is possible provided the NIR laser power is only weakly perturbing to the diamond's librational motion. 

Aside from the example given of magnetometry, a more general case to consider is how photon shot noise limits the minimum detectable deflection angle. Accurately calculating this limit is difficult due to the random shape and size of trapped microdiamonds often resulting in a variable trapping frequency and photon count rates for each particle, not to mention variable moments of inertia which scale as $r^5$. An estimated angular deviation of a few degrees in the 1\,ms of induced motion is a reasonable starting point based on the findings in the main text (Fig 4) and numerical calculation of the deflection from the equation of motion (Eq. \ref{eq:eom}). Thus, with a readout contrast of $C = 50$\% easily achieved in typical SMC Rabi measurements, the photon counts will vary between $x_{|0\rangle}$ for $\theta = \theta_0$ and $x_{|-1\rangle} = Cx_{|0\rangle}$ for $\theta_{-1}$ the shot-noise limited angular sensitivity is calculated from 
\begin{equation}
\delta S = \frac{x_{|0\rangle}(1-C)}{\theta_0 - \theta_{-1}}\delta \theta = \frac{1/2 x_{|0\rangle}}{4^\circ}\delta \theta.
\label{eq:ee}
\end{equation}
Assuming $\delta S = \sqrt{\frac{3x_{|0\rangle}}{4}}$ (halfway between the two angular extrema to ensure linear operation) we find
\begin{equation}
\delta \theta  = \frac{4^\circ\sqrt{3}}{\sqrt{x_{|0\rangle}}}\sqrt{t_d}
\label{eq:dd}
\end{equation}
With $x_{|0\rangle} \approx 10000$ in a single measurement as before, the angular sensitivity evaluates to 0.1\,mrad/$\sqrt{\text{Hz}}$.    

\subsection{Spin projection noise}
We now consider the case of noise sources other than detection issues that affect the ability to measure the rotation angle of the diamond. If the diamond were in a quantum superposition of orientations, measurements of the orientation after sufficient time for the spin torque to drive a significant rotation would yield two distinct values due to the two states the NV spin may be found in, and hence exhibit projection noise. However, for the case of $N$ uncorrelated spins, the diamond is not in a superposition of orientations, and so the particle position measurement is purely classical. Spin projection noise nevertheless contributes to the noise in the measured angle. Assuming operation at the mid-fringe point of the sequence, the spin ensemble following the microwave pulse sequence will be projected into $\approx N/2$ NVs in each spin state, with an uncertainty of $\pm\sqrt{N}$ stemming from spin projection noise~\cite{sdegen_quantum_2017}. The particle magnetisation, spin torque and ultimately particle rotation angle will therefore reflect the spin projection noise from each projective readout of the spin ensemble.

The torque noise from projection noise will be
\begin{equation}
\delta \tau = \hbar \gamma_e B \delta N = \hbar \gamma_e B\sqrt{N/2},
\end{equation}
with $\gamma_e/2\pi = 28\,$GHz/T and $B$ the magnetic bias field, assumed to be aligned for maximum spin torque. We perform measurements of the torque at a rate of $(t + t_d)^{-1}$ per second, with $t$ the measurement time (including the $T_2^\ast$-long Ramsey sequence and the time required for the particle to rotate to an angle we can detect) and $t_d$ the total time required to reset the particle's motion after each measurement (including optical pumping), typically $t_d \gg t$. The spin-projection torque noise is 
\begin{equation}
\delta \tau \approx \hbar \gamma_e B \delta N \sqrt{t_d}.
\end{equation}
For $N = 10^8$ spins, $t_d = 10\,$ms, $B = 27\,$mT, this is a projection noise limited torque sensitivity of $10^{-22}\,$N\,m/$\sqrt{\text{Hz}}$. This torque will drive the particle to rotate an angle
\begin{equation}
\delta \theta \approx \frac{\delta \tau}{I}f(t, \omega_0, \gamma_g, T_1)
\end{equation}
in time $t$ of spin-mechanical conversion. Here, $I$ is the moment of inertia and the function $f(t, ...)$ contains the explicit time dependence (\emph{eg.} damped relaxation or underdamped oscillations) of the angular time evolution due to gas damping $\gamma_g$, $T_1$ relaxation and the harmonic trapping frequency $\omega_0$.  The projection noise-limited minimum detectable angle in one second of averaging is 
\begin{equation}
\delta \theta =\frac{\hbar \gamma_e B\sqrt{N/2}}{I}f(t, \omega_0, \gamma_g, T_1)\sqrt{t_d}.
\end{equation}
With  $f(t, \omega_0, \gamma_g, T_1)\approx t^2/2$ (valid for short times $\ll T_1, 2\pi/\omega_0, \gamma_g^{-1}$ as in our current work), $N = 10^8$ spins, $t_d = 10\,$ms, $B = 27\,$mT and $I \approx 10^{-23}\,$kg\,m$^2$, the spin-projection noise limited angular measurement will be of order 0.1\,microradians\,/$\sqrt{\text{Hz}}$. For a given particle rotation angle, angular noise from spin projection noise will be $\delta \theta/\theta = \sqrt{2/N}\approx 10^{-4}$ for the high-NV density particles we use in the current experiment. 

\subsection{Thermal fluctuations}
The intrinsic thermal motion in the trap of the librating degree of freedom excited in SMC will contribute noise that limits the angular resolution. While a levitated, librating diamond acts as an archetypal torsion pendulum (one of the most sensitive types of weak force measurements), the contribution of thermal noise to our SMC measurement is analysed differently to the usual manner \cite{schen_thermal_1990}. In the most sensitive torsion pendulum experiments, the torque to be detected is applied at or near to the resonance frequency of the oscillator, which in the case of our experiment would be at the librational frequency $\omega_0/2\pi\sim 0.50$-$1\,$kHz. In this case, the minimum detectable torque is given by $\delta \tau = \sqrt{4 k_B T I \gamma_g /\delta t}$ ~\cite{shaiberger_highly_2007}, where $\delta t$ is the measurement time. Essentially, a cumulative response builds up due to application of the resonant torque, while thermal noise is suppressed as $1/\sqrt{\delta t}$. This mode of operation is best suited to resonators with high quality factors (low damping coefficients) and -- unique to our case -- long spin relaxation times. 

In the current realisation of our experiment, $\omega_0$, $\gamma_g$ and $T_1^{-1}$ are all roughly equal, so resonant operation as a torsion oscillator is not feasible. Instead of harmonic oscillations, we observe only the initial onset of an oscillation period before gas damping suppresses motion. Therefore, the oscillator is well equilibrated before each experiment with the background gas and the equipartition theorem yields 
\begin{equation}
\langle \theta \rangle = \sqrt{\frac{k_B T}{I \omega_0^2}},
\end{equation} 
\emph{i.e.} a single measurement of the diamond's orientation will have zero mean (relative to some arbitrary initial axis) and standard deviation $\langle \theta \rangle$. Performing a measurement at a rate $(t + t_d)^{-1}\approx (t_d)^{-1}$ improves the angular sensitivity to $\langle \theta \rangle \sqrt{t_d}$, which for $T = 300$\,K, $t_d = 10\,$ms, $I \approx 10^{-23}\,$kg\,m$^2$ and $\omega_0/2\pi = 1\,$kHz gives a thermal-noise limited angular sensitivity of $0.3\,$mrad/$\sqrt{\text{Hz}}$. Improvement of this sensitivity can be achieved via cooling of the librational mode temperature~\cite{sdelord_spin-cooling_2020}, increasing the trap frequency~\cite{sskakunenko_strong_2025} or reducing the mechanical reset time $t_d$. Entering the underdamped regime, where spin torques yield significantly greater angular deviations and harmonic oscillations rather than the overdamped, transient responses we witness in current experiments is an attractive prospect for improving the measurement sensitivity and requires pumping down the vacuum system. However, harmonic oscillations and higher $Q$-factors at reduced pressures result in long mechanical reset times and thus reduced duty cycle unless measures are taken to quickly cool or damp the induced motion to reset the oscillator for the next SMC measurement.

\section{Macroscopic quantum superpositions}
Here we provide some details and an outlook for observations of quantum mechanical behaviour of the particle's motion. A simple question to ask is: what happens when the NV spins are put into a coherent superposition, eg. by a microwave $\pi$/2-pulse, and whether this spin superposition has any effect on the mechanical state of the particle. Can the particle also be put into a superposition of orientation states? Does mechanical decoherence driven by air collisions affect spin coherence, or \emph{vice versa}? There are two regimes to consider. In the first, the particle is in a well-defined quantum state of motion and there is strong collective coupling between it and the spin ensemble. Here, spin dephasing results in mechanical dephasing and vice versa. The present work is far outside of this regime, given we operate at atmospheric pressures. In the second regime, that of an oscillator with large dissipation, spin coherence could potentially be affected by the spin-mechanical coupling. While we are insensitive to these effects \emph{a priori} in the present work, we also show here these effects to be small in the absence of quantum correlations within the spin ensemble or of a strong single-spin-particle coupling.\\

To illustrate this, let us consider the case of a spin ensemble initialised by optical polarization followed by a microwave field inducing a $\pi$/2 rotation of each spin. Importantly, the resulting state is not a correlated superposition state, but a product state of coherent superpositions. First taking a two-spin ensemble as an example, we have:
\begin{equation}
|\Psi\rangle_{spin} = \frac{1}{2}\,(|1\rangle + |0\rangle)_1 \otimes (|1\rangle + |0\rangle)_2
= \frac{1}{2}\,\bigl(|11\rangle + |00\rangle + |10\rangle + |01\rangle\bigr),
\end{equation}
which is itself a superposition of three Dicke states with 0, 1 and 2 spins in $m_s=1$. Following spin mechanical conversion, each spin up imparts to the diamond a certain angular momentum and small shift of its orientation $\theta$, the spins therefore become entangled with the orientation of the diamond particle, which we can write as:
\begin{equation}
|\Psi\rangle_{spin,m} = \frac{1}{2}\,\bigl(|11\rangle\otimes|2\theta\rangle_{\rm m} + |00\rangle\otimes|0\rangle_{\rm m} + (|10\rangle + |01\rangle)|\theta\rangle_{\rm m}\bigr).
\end{equation}
Note here that to generate a cat state, that is a macroscopic superposition of particle orientations, we desire the states $|11\rangle\otimes|2\theta\rangle_{\rm m} + |00\rangle\otimes|0\rangle_{\rm m}$, whereas in reality there are three distinct orientations the particle can be found in due to the two other degenerate Dicke states. In the current experiment, a single-spin induced perturbation $\theta$ is negligible and not measurable over the coherence time of the spins: we can only detect rotation induced by a large ensemble of spins. Generalising the above to an ensemble of $N$ spins, each initialised by optical polarization followed by a microwave $\pi/2$ pulse, the resulting state is a also product of coherent superpositions,

\begin{equation}
|\Psi\rangle_{\text{spin}}
= \bigotimes_{i=1}^{N} \frac{1}{\sqrt{2}}\left(|1\rangle_i + |0\rangle_i\right).
\end{equation}

We can rewrite this state in the symmetric Dicke basis. Let $|D_N^{(k)}\rangle$ denote the Dicke state with $k$ spins in $m_s=1$ and $(N-k)$ spins in $m_s=0$:

\begin{equation}
|D_N^{(k)}\rangle
= \frac{1}{\sqrt{\binom{N}{k}}}
\sum_{\text{perm}} | \underbrace{1\dots1}_{k}
\underbrace{0\dots0}_{N-k} \rangle,
\end{equation}
where we sum over all possible permutation of the spin ensemble. The spin state can then be written as

\begin{equation}
|\Psi\rangle_{\text{spin}}
= \frac{1}{2^{N/2}}
\sum_{k=0}^{N}
\sqrt{\binom{N}{k}}
\, |D_N^{(k)}\rangle.
\end{equation}

\begin{figure}[h]
  \centering
  \includegraphics[width=0.8\textwidth]{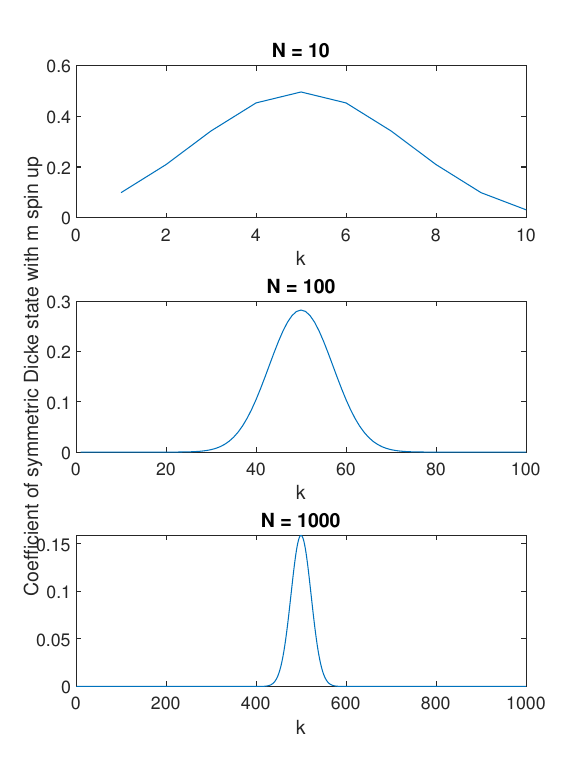}
  \caption{Weights of the Dicke states $|D_N^{(k)}\rangle$ with $k$ spins in $m_S = 1$ for three different numbers of total spins $N$.}
  \label{fig:dicke}
\end{figure}

The coherent spin state is thus a superposition of Dicke states with a binomial weight distribution peaked around $k=N/2$. The fully polarised states ($k=0$ and $k=N$), which constitute a GHZ-type state, carry exponentially small weights $\sim 2^{-N}$. Following spin–mechanical conversion, a Dicke state with $k$ spins in $m_s=1$ correlates with a mechanical orientation state $|k\theta\rangle_{\mathrm{m}}$ and the joint spin–mechanical state becomes:

\begin{equation}
|\Psi\rangle_{\text{spin,m}}
= \frac{1}{2^{N/2}}
\sum_{k=0}^{N}
\sqrt{\binom{N}{k}}
\, |D_N^{(k)}\rangle
\otimes |k\theta\rangle_{\mathrm{m}}.
\end{equation}

This state is entangled: each collective spin projection $k$ is correlated with a distinct mechanical orientation. However, for large numbers of spins $N$, the weights of the Dicke states tends to concentrate around its maximum at $k=N/2$, with the width of the distribution decreasing relative to $N$. Fig \ref{fig:dicke} shows numerical simulations for a small number of spins (up to $N=1000$) exhibiting this behaviour. Such an effect, in the current experimental conditions where only a large amount of spins induces a measurable rotation of the diamond will strongly suppress the impact of mechanical decoherence on the spin properties. While much more complicated from an experimental point of view, entangling the motional state of the particle with that of a single NV spin offers a conceptually more straightforward approach to generating a macroscopic superposition state. 

\section{MDMR vector magnetometry} 
\begin{figure*}
	\centering
		\includegraphics[width = \textwidth]{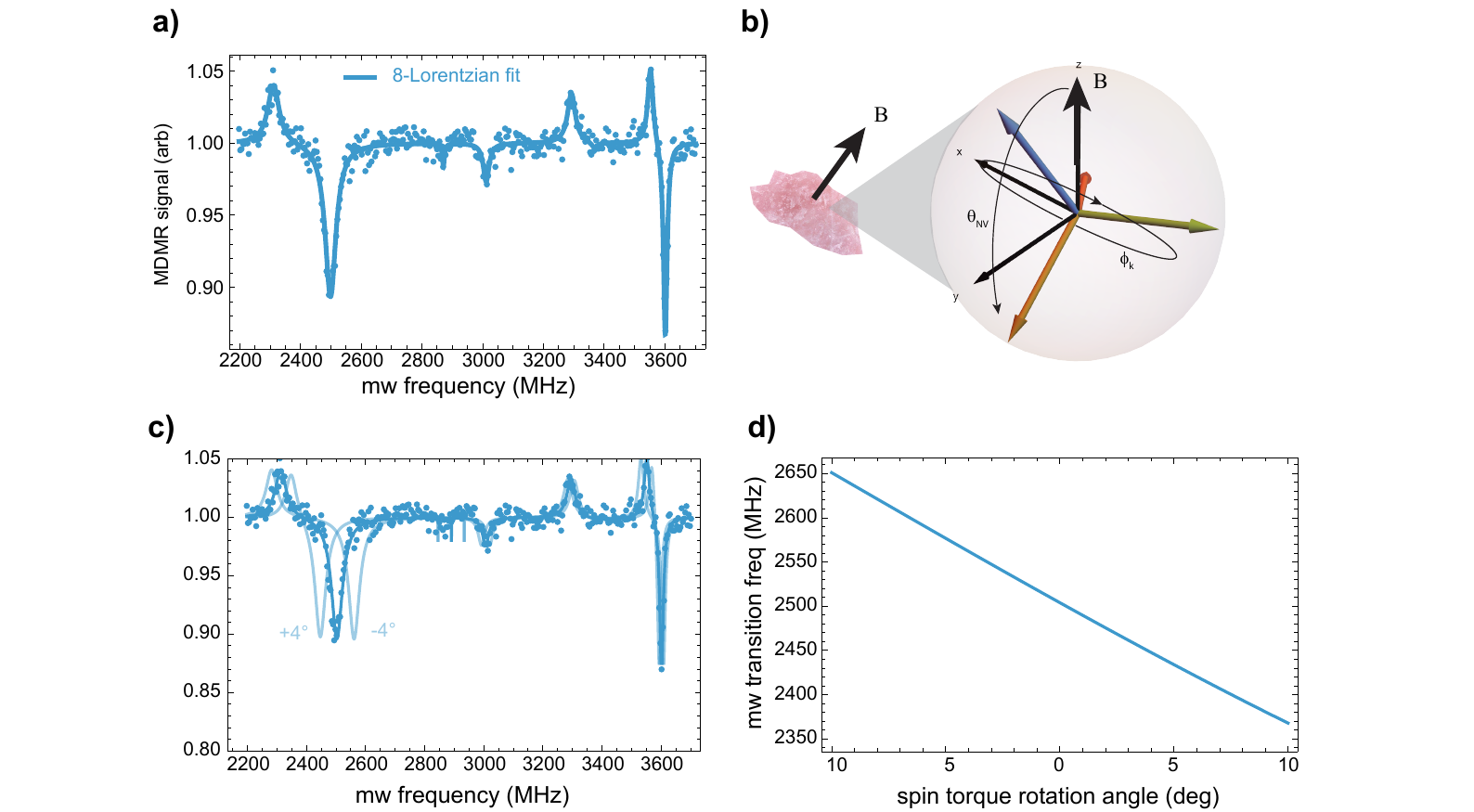}
	\caption{MDMR vector magnetometry to determine angular displacement. a) cw-MDMR spectrum and fit for the microdiamond particle used in experiments (main text, Fig 4). b) The microdiamond is trapped at an arbitrary orientation with respect to the magnetic field. The MDMR spectrum is thus referenced to the direction of the applied magnetic field $\boldsymbol{B}$ (here along the $z$-axis). Two angles, $\theta_\text{NV}$ and $\phi_k$ parametrise the NV crystallographic orientation with respect to the applied magnetic field. c) cw-MDMR data overlaid with best-fit MDMR spectrum (bold) corresponding to $|\boldsymbol{B}|$ = 271.5\,G, $\theta_\text{NV} = 225\,^\circ$, $\phi_k = 292.98^\circ$. Fainter traces depict spectrum resulting from $\pm4\,^\circ$ angular shifts arising from spin torques applied to align the NV orientation class targeted with microwaves (orange in (b)) to the magnetic field. d) linearised rotation angle to transition frequency calibration.}
	\label{fig:figs4}
\end{figure*}

We measure the angle the microdiamond rotates after application of a $\pi$-pulse using the pump-probe microwave pulses described in Fig. 4 of the main text. To quantify the rotation angle, we use a model of the MDMR spectrum to estimate the variation of the MDMR transition frequency as a function of rotation angle for the NV orientation class we target with microwaves. Fig. \ref{fig:figs4}(a) depicts the steady-state MDMR spectrum obtained with continuous application of light and microwaves, and an overlaid Lorentzian fit to extract the eight transition frequencies. Next, we construct the NV spin Hamiltonian
\begin{equation}
\label{Eq:hnv}
H=D_{\text{zfs~}}\hat{S}_\text{z}^2+\gamma_{e}\vec{\mathbf{B}}\cdot \vec{\mathbf{S}},
\end{equation}
where $D_{\text{zfs~}}/2\pi = 2870\,$MHz is the zero-field splitting, $\gamma_{e}/2\pi=2.803$~MHz/G is the electron gyromagnetic ratio and $\vec{\mathbf{S}}=(\hat{S}_x,\hat{S}_y,\hat{S}_z)$ are the spin-1 operators. Defining a magnetic field $\boldsymbol{B}$ and angles $\theta_\text{NV}$ and $\phi_k$ as shown in Fig. \ref{fig:figs4}(b), we compute the resulting electron spin transition frequencies and use numerical least-squares minimisation to determine the $\boldsymbol{B}$, $\theta_\text{NV}$ and $\phi_k$ that give rise to the MDMR spectrum we see in Fig. \ref{fig:figs4}(a). The MDMR spectrum $f$ for the optimum values $|\boldsymbol{B}|$ = 271.5\,G, $\theta_\text{NV} = 225\,^\circ$, $\phi_k = 292.98^\circ$ is shown in Fig. \ref{fig:figs4}(b) using the fitted amplitudes and widths from Fig. \ref{fig:figs4}(a). Next, we assume that the spin torque when the resonant microwaves populate the $m_S = -1$ spin state acts to align the relevant NV orientation class (at $2500\,$MHz) towards the magnetic field. Using this model, we obtain in Fig. \ref{fig:figs4}(d) a plot of $f(\theta_d)$, which informs the frequency shift imparted by a spin torque mediated rotation towards the magnetic field. Inverting the frequency vs. angle data allows us to then calibrate the frequency measurements determined in Fig. 4 of the main text into rotation angles.

The torque is related to the angular momentum of the particle by $\boldsymbol{\tau} = \frac{d\boldsymbol{L}}{dt} = I\frac{d\boldsymbol{\omega}}{dt}$, with $I$ the moment of inertia. Fitting a quadratic function to the data in Fig 4(c) of the main text yields $\theta(t) = 2.18\times 10^6 \text{rad}\,\text{s}^{-2} t^2$. Differentiating twice with respect to time yields $\frac{d\omega}{dt} = 4.35\times10^6\,\text{rad}\,\text{s}^{-2}$. We approximate the particle moment of inertia as an average between a sphere of radius $r$ and a cube of side length $2\,r$, yielding $I_\text{eff} = 1/2\left(1/6 m r^2 + 2/5 m r^2\right)$, with $m = 1.8\,$pg for a $r =5\,\upmu$m particle. Thus, we find $\tau = I_\text{eff} \frac{d\omega}{dt} = 56.5\,$aN\,m. The torque applied per spin is approximated by $\tau_s = \hbar \gamma B$, implying $1\times 10^8$ spins are flipped by the initial $\pi$-pulse.

\section*{References}

\end{document}